\documentclass[aps,prd,reprint,superscriptaddress,nofootinbib,longbibliography]{revtex4-2}

\usepackage[utf8]{inputenc}
\usepackage[T1]{fontenc}
\usepackage{amsmath,amssymb,bm,mathtools}
\usepackage{graphicx}
\usepackage{hyperref}
\hypersetup{hidelinks}
\usepackage{microtype}

\begin{document}

\title{Persistence of measurement-induced nonlocality in uniformly accelerating Unruh–DeWitt detectors}

\author{Shi-Pu Gu}
\affiliation{College of Electronic and Optical Engineering and College of Flexible Electronics (Future Technology), Nanjing University of Posts and Telecommunications, Nanjing 210023, China}
\author{Ming-Ming Du}
\email{mingmingdu@njupt.edu.cn}
\affiliation{College of Electronic and Optical Engineering and College of Flexible Electronics (Future Technology), Nanjing University of Posts and Telecommunications, Nanjing 210023, China}






\author{Yu-Bo Sheng}
\affiliation{School of Physics, Hangzhou Normal University, Hangzhou, 311121, China}

\author{Lan Zhou}
\affiliation{School of Physics, Hangzhou Normal University, Hangzhou, 311121, China}

\date{October 18, 2025}

\begin{abstract}
Uniform acceleration induces a thermal response to the Minkowski
vacuum and can thereby modify quantum correlations. For measurement-induced nonlocality (MIN), a form of nonclassical correlation distinct from entanglement, previous field-mode analyses of bosonic fields found that it decreases with increasing acceleration and vanishes in the infinite-acceleration limit. This raises the question of whether the disappearance of MIN is a generic consequence of the Unruh effect or depends on the physical description of the accelerated quantum system. To address this question, we study two uniformly accelerating Unruh-DeWitt detectors interacting with a massless scalar field. We find that the response of MIN to the Unruh temperature depends sensitively on the initial detector state: it can decrease monotonically, vanish at an intermediate temperature and subsequently recover, or increase monotonically. Moreover, for a broad class of initial states, MIN approaches a nonzero value in the high-temperature limit. These results show that the suppression of MIN under strong acceleration is not universal. Instead, Unruh-induced detector dynamics gives rise to a state-dependent response in which measurement-induced nonlocality can be suppressed, restored, or enhanced.

\end{abstract}

\maketitle

\section{Introduction}
\label{sec:introduction}
A uniformly accelerated detector coupled to a quantum field in the
Minkowski vacuum exhibits a thermal response characterized by the
Unruh temperature $T_{\mathrm U}=a/(2\pi)$ in natural units
\cite{Unruh1976,Crispino2008}. The Unruh effect therefore provides
a natural setting for studying how acceleration modifies quantum
correlations in relativistic quantum systems~\cite{PeresTerno2004,Alsing2012}. Early studies of quantum information in noninertial frames showed
that acceleration can degrade entanglement in both bosonic and
fermionic systems
~\cite{Alsing2003,Fuentes2005,Alsing2006,Adesso2007}
and reduce the fidelity of quantum-information protocols such as
teleportation~\cite{Alsing2003,Landulfo2009}. However, the extent and even the qualitative form of
this degradation depend on both the correlation measure and the
physical description of the accelerated system~\cite{Datta2009,Bruschi2010,Bhuvaneswari2022}. The response of
nonclassical correlations to the Unruh effect therefore cannot, in general, be inferred from entanglement alone.

The response of quantum correlations to acceleration is, however,
strongly measure dependent. Studies of accelerated quantum systems have shown that correlations
beyond entanglement can remain nonzero even when entanglement is
strongly degraded
~\cite{Datta2009,Wang2010,Wang2016,Bhuvaneswari2022,Li2024MSC,Bachain2026},
while some correlation measures exhibit nonmonotonic behavior or
revival as the acceleration or effective temperature is varied
~\cite{Bhuvaneswari2022,Li2024MSC,Wang2025,Wang2025Coherence,Wu2025Coherence}. Such
differences arise because distinct correlation measures characterize
inequivalent nonclassical features of a quantum state and therefore
need not respond in the same way to acceleration-induced thermal
effects. Increasing the Unruh temperature thus does not necessarily
lead to a monotonic suppression of all forms of quantum correlation.
It is therefore important to examine different correlation measures
individually rather than infer their relativistic behavior from
entanglement alone.

A natural quantity to examine in this regard is
measurement-induced nonlocality (MIN). Introduced by Luo and Fu~\cite{Luo2011},
MIN quantifies the maximal global disturbance of a bipartite state
induced by local projective measurements that leave the reduced state
of the measured subsystem unchanged. Despite its
name, MIN is distinct from Bell nonlocality and characterizes a
different aspect of nonclassical correlations. Its behavior under
acceleration was studied by Tian and Jing~\cite{Tian2013} for Dirac and bosonic
fields beyond the single-mode approximation. They
found qualitatively different asymptotic behaviors: MIN associated
with Dirac fields remains finite at arbitrary acceleration, whereas
for bosonic field modes it decreases and vanishes in the
infinite-acceleration limit~\cite{Tian2013}. Within that field-mode
framework, MIN is therefore completely lost at sufficiently large
acceleration. Whether this disappearance is a generic consequence of
the Unruh effect or instead depends on the physical description of
the accelerated quantum system remains unclear.

A natural way to address this question is to consider localized
quantum probes interacting with the field. The Unruh--DeWitt (UdW)
detector provides a standard framework for this purpose, modeling a
localized quantum system coupled to a quantum field along a prescribed
spacetime trajectory~\cite{Takagi1986,Hu2012,Lee2014,Brown2013}. It has been
widely used to study the response of accelerated quantum systems and
the evolution of correlations generated through detector-field
interactions~\cite{LinHu2010,Hu2012,Brown2013,Lee2014,Salton2015,PozasKerstjens2015,Li2024MSC,Wang2025Coherence,Du2026}. In this framework, the influence of acceleration is
encoded in the field correlation functions sampled along the detector
trajectory, providing a description that is physically distinct from
analyses based directly on Minkowski and Rindler field modes. This
makes the UdW model particularly suitable for examining how
correlations initially encoded between localized subsystems respond
to the Unruh effect. The relevant question is therefore whether the
suppression of MIN found in bosonic field-mode treatments persists in
the detector setting, or whether its response depends on the initial
detector state and the corresponding detector--field dynamics.

In this work, we investigate measurement-induced nonlocality in two
uniformly accelerating Unruh--DeWitt detectors coupled to a massless
scalar field. We examine how the stationary MIN depends on the Unruh
temperature and on the initial-state structure of the detector pair.
We find that its temperature dependence is strongly state dependent.
Depending on the initial state, MIN can exhibit monotonic suppression,
complete suppression followed by recovery, or monotonic enhancement.
For a broad class of initial states, MIN approaches a nonzero
asymptotic value in the high-temperature limit. This behavior
contrasts with the vanishing MIN previously obtained for bosonic field
modes in the infinite-acceleration limit~\cite{Tian2013}. The resulting
picture is therefore not one of universal degradation: acceleration
can suppress, restore, or enhance MIN depending on the initial
detector state.

The remainder of this paper is organized as follows.
In Sec.~II, we review the definition of MIN and its expression
for two-qubit states. In Sec.~III, we introduce the model of two
uniformly accelerating Unruh--DeWitt detectors interacting with
a massless scalar field and derive the reduced detector dynamics.
In Sec.~IV, we analyze the dependence of MIN on the Unruh
temperature and the initial-state parameter, with particular
attention to its high-temperature behavior. Finally, Sec.~V
summarizes our main results.

\section{Measurement-Induced Nonlocality}\label{sec:MIN}
Measurement-induced nonlocality (MIN)~\cite{luo2011a}
quantifies the global effect of locally invariant measurements
on a bipartite quantum state. Given a bipartite state
$\rho_{AB}$, MIN is defined as~\cite{luo2011a}
\begin{equation}
N(\rho_{AB})=\max_{\Pi^A}
\left\|\rho_{AB}-\Pi^A(\rho_{AB})\right\|^2,
\label{eq:min-definition}
\end{equation}
where $\Pi^A=\{\Pi_k^A\}$ denotes a set of local projective
measurements on subsystem $A$,
$\Pi^A(\rho_{AB})
=
\sum_k
(\Pi_k^A\otimes I_B)
\rho_{AB}
(\Pi_k^A\otimes I_B)$,
and
$\|X\|^2=\operatorname{Tr}(X^\dagger X)$ denotes the squared
Hilbert--Schmidt norm~\cite{bhatia1997}.
The maximization is taken over all local projective measurements
that leave the reduced state
$\rho_A=\operatorname{Tr}_B(\rho_{AB})$ invariant, i.e.,
$\sum_k\Pi_k^A\rho_A\Pi_k^A=\rho_A$.
MIN thus characterizes the global disturbance induced by a
locally invariant measurement on one subsystem.

For a two-qubit state $\rho$ with Bloch representation,
\begin{align}
\rho={}&\frac{1}{2}\frac{I}{\sqrt{2}}\otimes\frac{I}{\sqrt{2}}
+\sum_{i=1}^{3}x_i X_i\otimes\frac{I}{\sqrt{2}}
+\frac{I}{\sqrt{2}}\otimes\sum_{j=1}^{3}y_jY_j \notag\\
&+\sum_{i=1}^{3}\sum_{j=1}^{3}t_{ij}X_i\otimes Y_j,
\label{eq:bloch}
\end{align}
where $X_i=\sigma_i/\sqrt{2}$ and
$Y_j=\sigma_j/\sqrt{2}$, with $\sigma_i$ denoting the Pauli
matrices. The coefficients are determined by the
Hilbert--Schmidt inner product as
\begin{equation}
\begin{aligned}
x_i&=\operatorname{Tr}\left[
\rho\left(X_i\otimes\frac{I}{\sqrt{2}}\right)\right],\\
y_j&=\operatorname{Tr}\left[
\rho\left(\frac{I}{\sqrt{2}}\otimes Y_j\right)\right],\\
t_{ij}&=\operatorname{Tr}\left[\rho(X_i\otimes Y_j)\right].
\end{aligned}
\label{eq:coefficients}
\end{equation}

Define the vectors
$\bm{x}=(x_1,x_2,x_3)^\top$,
$\bm{y}=(y_1,y_2,y_3)^\top$,
and the correlation matrix $\mathrm{T}=(t_{ij})$.
Let $M=\mathrm{T}\mathrm{T}^\top$ and
$\|\bm{x}\|^2=\sum_i x_i^2$.
The matrix $M$ is real, symmetric, and positive semidefinite.
According to Ref.~\cite{luo2011a}, MIN for a general two-qubit
state is given by
\begin{equation}
N(\rho)=
\begin{cases}
\operatorname{Tr}(M)
-\dfrac{\bm{x}^\top M\bm{x}}{\|\bm{x}\|^2},
& \bm{x}\neq 0,\\[6pt]
\operatorname{Tr}(M)-\lambda_3,
& \bm{x}=0,
\end{cases}
\label{eq:twoqubit-min}
\end{equation}
where $\lambda_3$ denotes the smallest eigenvalue of
$M=\mathrm{T}\mathrm{T}^\top$.

\section{Unruh-DeWitt Detector Models}\label{sec:model}
We consider two uniformly accelerating Unruh--DeWitt detectors,
modeled as identical two-level atoms locally coupled to a massless
scalar field in $(3+1)$-dimensional Minkowski spacetime
~\cite{benatti2004}. The total Hamiltonian of the detectors and the
field is
\begin{equation}
H=\frac{\omega}{2}\Sigma_3+H_\Phi+\mu H_I,
\label{eq:H}
\end{equation}
where $\omega$ denotes the energy-level spacing of each detector and $\Sigma_3\equiv
\sigma_3^{(A)}\otimes I^{(B)}
+
I^{(A)}\otimes\sigma_3^{(B)}$.
Here, $\sigma_3^{(\alpha)}$ denotes the Pauli operator associated
with detector $\alpha=A,B$, $H_\Phi$ is the Hamiltonian of the free
massless scalar field $\Phi(t,\bm{x})$, and $\mu$ denotes the
detector--field coupling strength. The interaction Hamiltonian is
given by
\begin{equation}
H_I=
\left(\sigma_2^{(A)}\otimes 1^{(B)}\right)\Phi(t,\bm{x}_1)
+
\left(1^{(A)}\otimes\sigma_2^{(B)}\right)\Phi(t,\bm{x}_2).
\label{eq:HI}
\end{equation}

We assume an initially factorized detector-field state,
$\rho_{\mathrm{tot}}(0)
=
\rho_{AB}(0)\otimes|0\rangle\langle0|$,
where $\rho_{AB}(0)$ denotes the initial state of the two detectors
and $|0\rangle$ is the field vacuum. The total detector--field system
then evolves unitarily. We further assume weak detector--field
coupling, characterized by the dimensionless parameter $\mu\ll1$.
Under the Markovian approximation, for which the field correlation
time is much shorter than the characteristic dynamical timescale of
the detectors, the reduced density matrix $\rho_{AB}(\tau)$ obeys a
Kossakowski--Lindblad master equation with respect to the proper time
$\tau$~\cite{gorini1976},
\begin{equation}
\frac{\partial\rho_{AB}(\tau)}{\partial\tau}
=
-i[H_{\mathrm{eff}},\rho_{AB}(\tau)]
+\mathcal{L}[\rho_{AB}(\tau)].
\label{eq:master}
\end{equation}
The dissipative contribution is
\begin{equation}
\mathcal{L}[\rho_{AB}]
=
\sum_{\substack{i,j=1,2,3\\\alpha,\beta=A,B}}
\frac{C_{ij}}{2}
\left[
2\sigma_j^{(\beta)}\rho_{AB}\sigma_i^{(\alpha)}
-
\left\{
\sigma_i^{(\alpha)}\sigma_j^{(\beta)},
\rho_{AB}
\right\}
\right].
\label{eq:dissipator}
\end{equation}
The coefficients $C_{ij}$ are determined by the field correlation
functions. Introducing the Wightman function
$G^+(x,x')=\langle0|\Phi(x)\Phi(x')|0\rangle$
and its Fourier transform,
\begin{equation}
G(\lambda)
=
\int_{-\infty}^{\infty}d\tau\,
e^{i\lambda\tau}G^+(\tau)
=
\int_{-\infty}^{\infty}d\tau\,
e^{i\lambda\tau}
\langle\Phi(\tau)\Phi(0)\rangle,
\label{eq:spectral}
\end{equation}
the coefficients $C_{ij}$ take the form
\begin{equation}
C_{ij}
=
\frac{\gamma_+}{2}\delta_{ij}
-i\frac{\gamma_-}{2}\epsilon_{ijk}\delta_{3,k}
+\gamma_0\delta_{3,i}\delta_{3,j},
\label{eq:kossakowski}
\end{equation}
where
\begin{equation}
\gamma_\pm
=
G(\omega)\pm G(-\omega),
\qquad
\gamma_0
=
G(0)-\gamma_+/2.
\label{eq:gammas}
\end{equation}

The effective Hamiltonian
$H_{\mathrm{eff}}=\frac{1}{2}\bar\omega\sigma_3$
contains the Lamb-shift correction, with the renormalized frequency
\begin{equation*}
\bar\omega
=
\omega+i[K(-\omega)-K(\omega)],
\end{equation*}
where
\begin{equation*}
K(\lambda)
=
\frac{1}{i\pi}\,
P\!\int_{-\infty}^{\infty}
d\omega'\,
\frac{G(\omega')}{\omega'-\lambda}
\end{equation*}
is the Hilbert transform.

For uniformly accelerating detectors with proper acceleration $a$,
the Wightman function satisfies the Kubo--Martin--Schwinger (KMS)
condition,
\begin{equation*}
G^+(\tau)=G^+(\tau+i\beta),
\end{equation*}
where $\beta=2\pi/a=1/T$ and $T=a/(2\pi)$ is the Unruh
temperature. The KMS condition implies the detailed-balance relation
\begin{equation}
G(\lambda)=e^{\beta\lambda}G(-\lambda).
\label{eq:kms}
\end{equation}
Using this relation, the coefficients $\gamma_\pm$ become
\begin{equation}
\gamma_+
=
\left(1+e^{-\beta\omega}\right)G(\omega),
\label{eq:gamma-plus}
\end{equation}
and
\begin{equation}
\gamma_-
=
\left(1-e^{-\beta\omega}\right)G(\omega).
\label{eq:gamma-minus}
\end{equation}
It is convenient to introduce the ratio
\begin{equation}
\gamma
\equiv
\frac{\gamma_-}{\gamma_+}
=
\frac{1-e^{-\beta\omega}}
{1+e^{-\beta\omega}}
=
\tanh(\beta\omega/2).
\label{eq:gamma-ratio}
\end{equation}
Thus, $\gamma$ depends on the Unruh temperature and the detector
energy gap through the KMS relation.
In the long-time limit, the master equation yields the stationary
reduced state of the two detectors. In the basis
$\{|00\rangle,|01\rangle,|10\rangle,|11\rangle\}$,
the stationary state takes the X form~\cite{benatti2004},
\begin{equation}
\rho_{AB}=
\begin{pmatrix}
A&0&0&0\\
0&C&D&0\\
0&D&C&0\\
0&0&0&B
\end{pmatrix},
\label{eq:rho-x}
\end{equation}
where the matrix elements are determined by $\gamma$ and the
initial-state parameter $\Delta_0$:
\begin{equation}
A=
\frac{(3+\Delta_0)(\gamma-1)^2}
{4(3+\gamma^2)},
\qquad
B=
\frac{(3+\Delta_0)(\gamma+1)^2}
{4(3+\gamma^2)},
\label{eq:AB}
\end{equation}
\begin{equation}
C=
\frac{3-\Delta_0-(\Delta_0+1)\gamma^2}
{4(3+\gamma^2)},
\qquad
D=
\frac{\Delta_0-\gamma^2}
{2(3+\gamma^2)}.
\label{eq:CD}
\end{equation}
Here,
$\Delta_0=
\sum_{i=1}^{3}
\operatorname{Tr}\!\left[
\rho_{AB}(0)\,
\sigma_i^{(A)}\otimes\sigma_i^{(B)}
\right]$,
is an initial-state parameter satisfying
$-3\leq\Delta_0\leq1$ for physical states
~\cite{benatti2004}. The dependence of the stationary state on the
initial detector state is therefore encoded in $\Delta_0$.

\section{MIN under the Unruh-DeWitt Detector Model}\label{result}
In this section, we analyze MIN for the stationary state of two
uniformly accelerating Unruh-DeWitt detectors.
Using Eqs.~\eqref{eq:coefficients} and \eqref{eq:rho-x}, we obtain
$\bm{x}=(0,0,(A-B)/2)^\top$,
$\bm{y}=(0,0,(A-B)/2)^\top$, and
$\mathrm{T}=\operatorname{diag}\!\left(D,D,(A+B-2C)/2\right)$.
Substituting these quantities into Eq.~\eqref{eq:twoqubit-min}
yields
\begin{equation}
N(\rho_{AB})=
\frac{\left[\Delta_0-\tanh^2\!\left(\frac{\omega}{2T}\right)\right]^2}
{2\left[3+\tanh^2\!\left(\frac{\omega}{2T}\right)\right]^2}.
\label{eq:min-final}
\end{equation}
The dependence of $N(\rho_{AB})$ on $T$ is entirely governed by the hyperbolic function $\tanh(\omega/2T)$, which encodes the effect of thermal noise perceived by the accelerating detectors. In the low-temperature limit $T\to0$, one has $\tanh(\omega/2T)\to1$, and the MIN approaches a constant value
\begin{equation}
N(\rho_{AB})\big|_{T\to0}=\frac{(\Delta_0-1)^2}{32}.
\label{eq:lowT}
\end{equation}
This corresponds to a nearly pure initial configuration where the influence of the thermal field is negligible, and the nonlocality is primarily determined by the intrinsic correlations of the detectors, represented by $\Delta_0$. In contrast, in the high-temperature (or large-acceleration) limit $T\to\infty$, $\tanh(\omega/2T)\to0$, giving
\begin{equation}
N(\rho_{AB})\big|_{T\to\infty}=\frac{\Delta_0^2}{18}.
\label{eq:highT}
\end{equation}
Although the Unruh radiation introduces strong decoherence, the MIN remains finite. This indicates that a portion of measurement-induced quantum correlation survives even when entanglement may have completely vanished (Shown in Ref.~\cite{Bhuvaneswari2022}).

\begin{figure}[t]
\centering
\includegraphics[width=0.88\columnwidth]{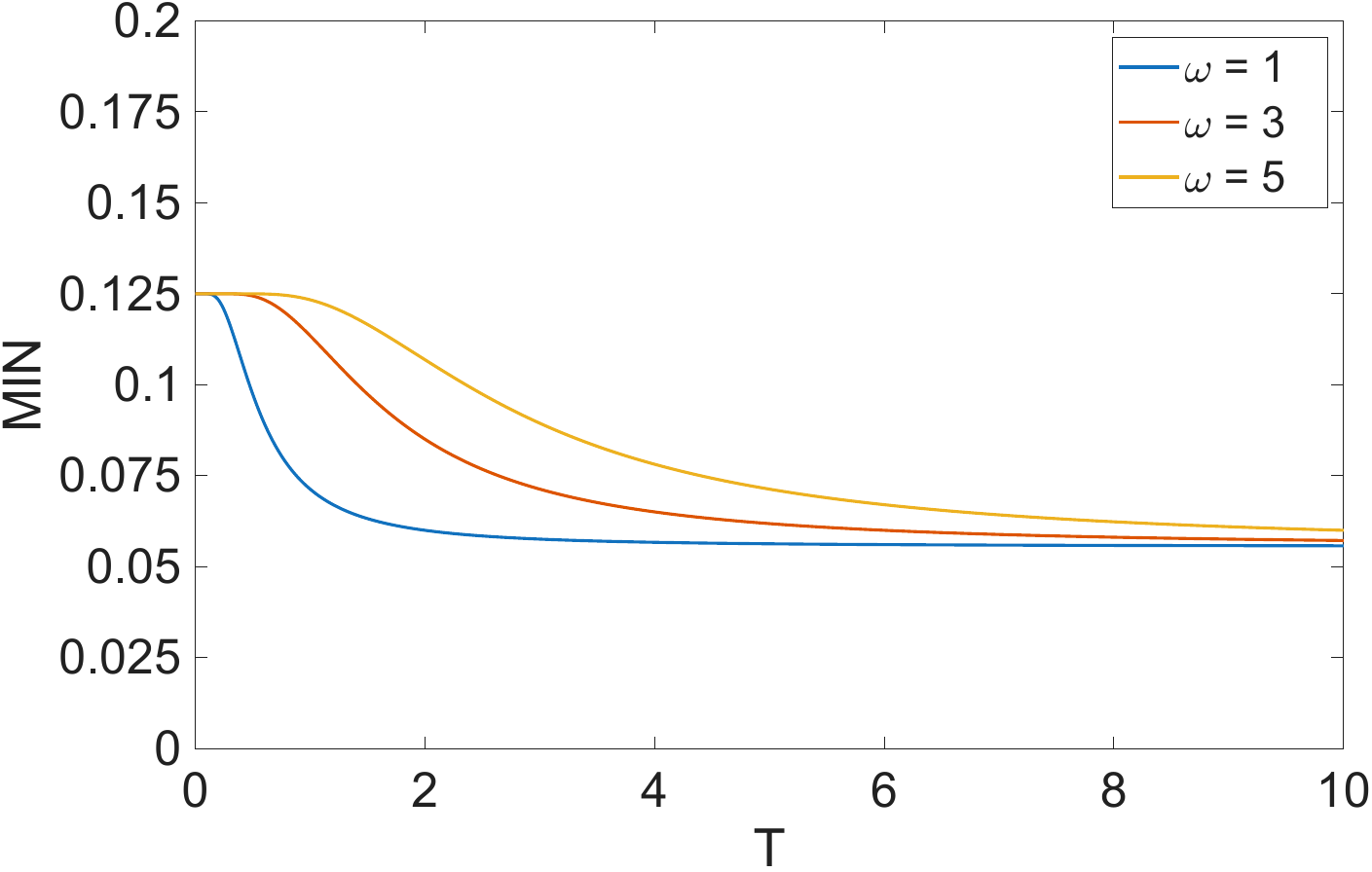}
\caption{MIN as a function of Unruh temperature for $\Delta_0=-1$ with energy gaps $\omega=1,3,5$.}
\label{fig:delta-minus-one}
\end{figure}

\begin{figure}[t]
\centering
\includegraphics[width=0.88\columnwidth]{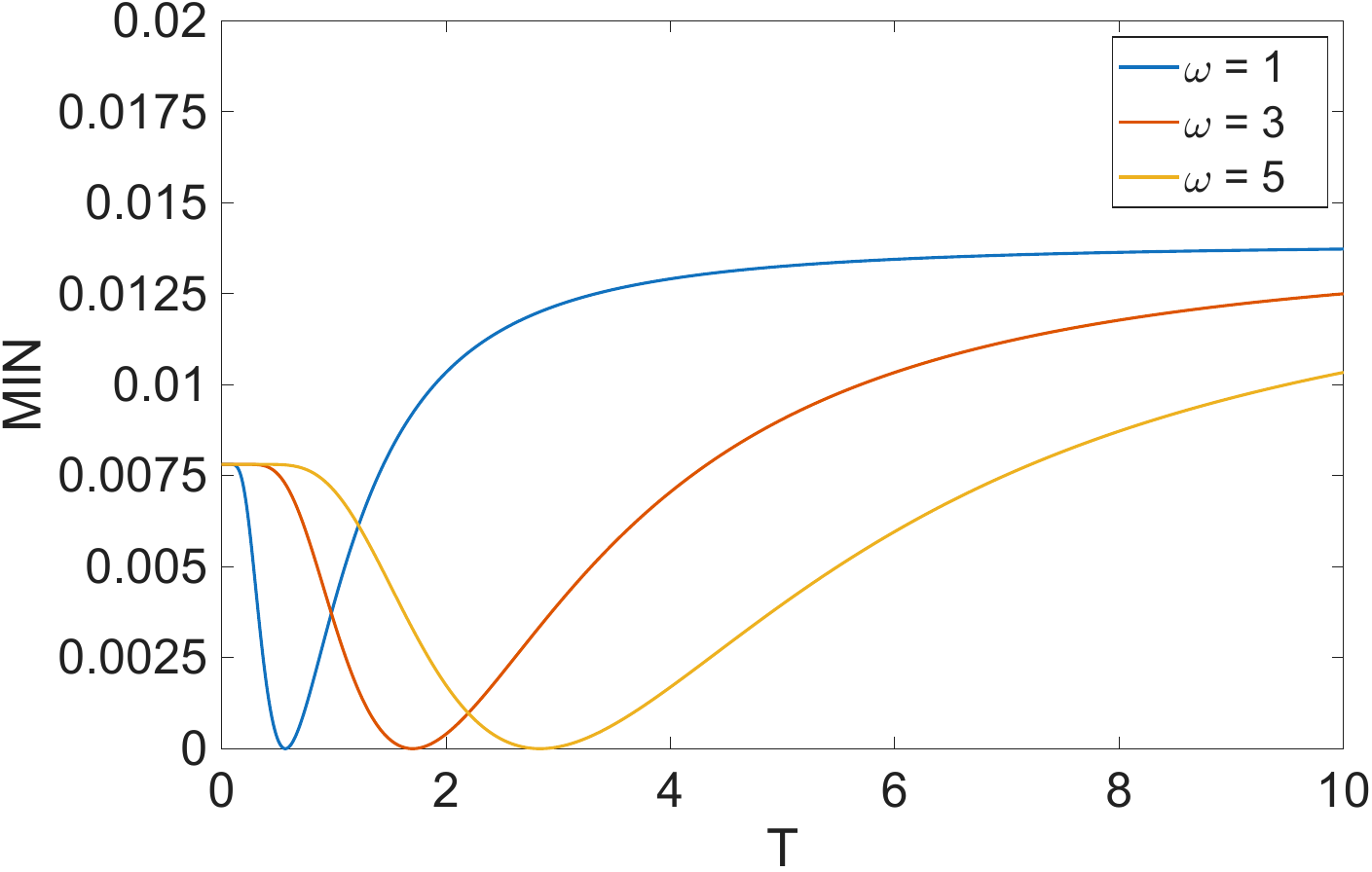}
\caption{MIN as a function of Unruh temperature for $\Delta_0=0.5$ with energy gaps $\omega=1,3,5$.}
\label{fig:delta-half}
\end{figure}

\begin{figure}[t]
\centering
\includegraphics[width=0.88\columnwidth]{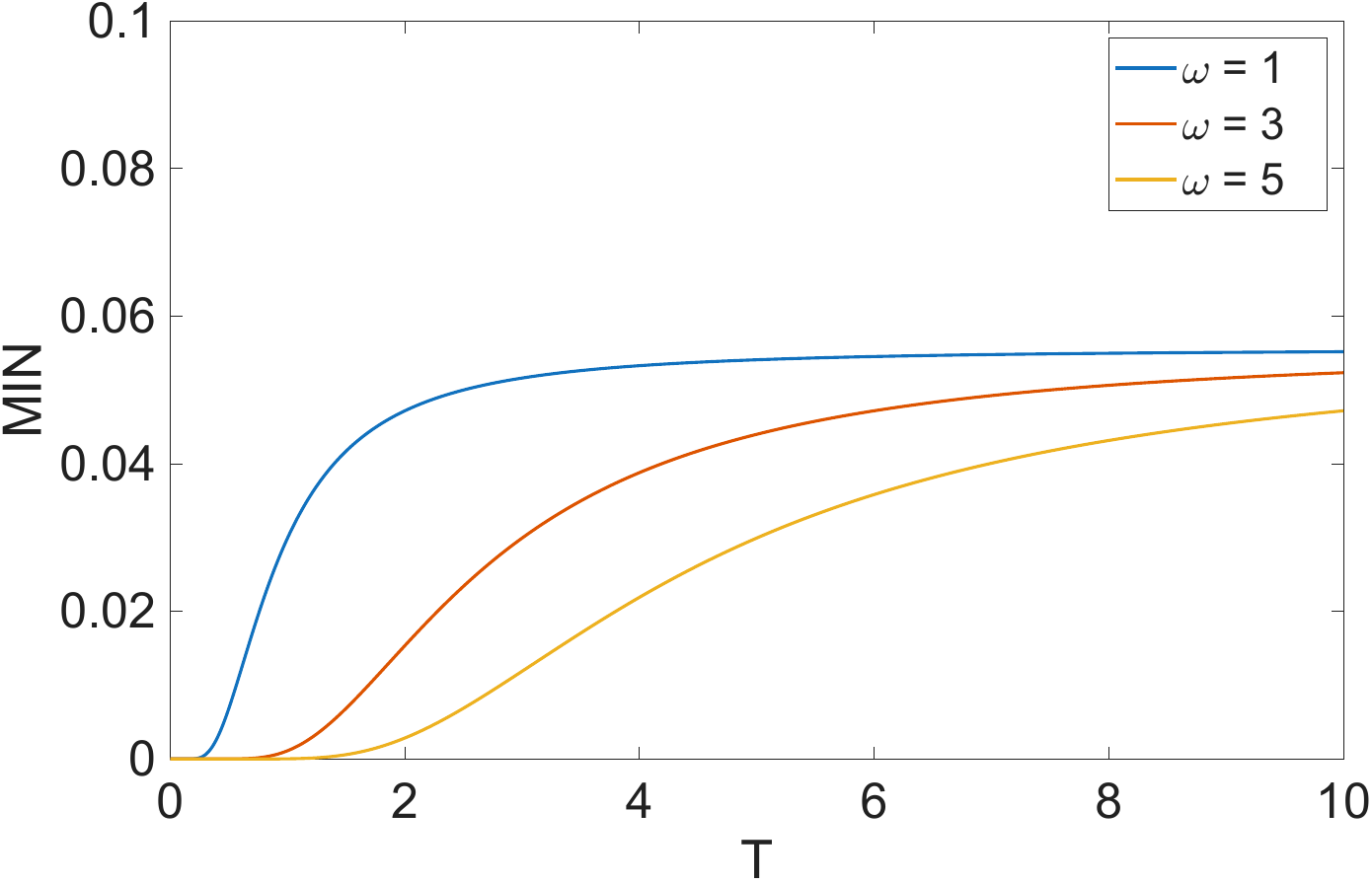}
\caption{MIN as a function of Unruh temperature for $\Delta_0=1$ with energy gaps $\omega=1,3,5$.}
\label{fig:delta-one}
\end{figure}

Moreover, Eq.~\eqref{eq:min-final} reveals that $N(\rho_{AB})$ vanishes only when $\Delta_0=\tanh^2(\omega/2T)$, implying that the equilibrium between the initial state correlations and the effective Unruh thermal parameter completely suppresses nonlocality. Figs.~\ref{fig:delta-minus-one}--\ref{fig:delta-one} illustrate the characteristic dependence of the measurement-induced nonlocality $N(\rho_{AB})$ on the Unruh temperature $T$ for representative values of the initial correlation parameter $\Delta_0$ and detector energy gaps $\omega$. The three plots serve as typical examples demonstrating the qualitative trends predicted by Eq.~\eqref{eq:min-final}. When $\Delta_0<0$, the MIN decreases monotonically with increasing Unruh temperature for all $\omega$, as shown in Fig.~\ref{fig:delta-minus-one}. In this regime, the initial correlations are anticorrelated, and the Unruh thermal noise continuously destroys quantum correlations, leading to a smooth and irreversible decay of $N(\rho_{AB})$. For intermediate values $0<\Delta_0<1$, the behavior becomes nonmonotonic, as depicted in Fig.~\ref{fig:delta-half}. Here, $N(\rho_{AB})$ first decreases to a minimum and then increases again at higher temperatures. This turning point occurs at $\tanh^2(\omega/2T)=\Delta_0$, where the analytical expression in Eq.~\eqref{eq:min-final} predicts the vanishing of MIN. The nonmonotonicity reflects the competition between the degradation of correlations due to Unruh thermalization and the residual nonlocal disturbance arising from measurement backaction. For the maximally correlated case $\Delta_0=1$, Fig.~\ref{fig:delta-one} shows that the MIN increases monotonically with temperature. At $T\to0$, MIN vanishes because $\tanh^2(\omega/2T)=1$ exactly cancels $\Delta_0$. As temperature increases, the mismatch between $\Delta_0$ and $\tanh^2(\omega/2T)$ grows, producing a monotonic increase in $N(\rho_{AB})$. This indicates that Unruh-induced thermal fluctuations can, paradoxically, enhance measurement-induced nonlocality in highly correlated initial states.

To further elucidate the role of the Unruh temperature, it is instructive to compare the behavior of MIN with other quantum correlation measures such as entanglement. For the same Unruh-DeWitt detector setup, entanglement typically exhibits a phenomenon known as ``sudden death'' (Shown in Ref.~\cite{Bhuvaneswari2022}): as the temperature increases beyond a critical value, the entanglement between the detectors abruptly vanishes and never revives. In contrast, the measurement-induced nonlocality $N(\rho_{AB})$ in Eq.~\eqref{eq:min-final} decays smoothly and remains finite even in the asymptotic high-temperature limit. This persistence indicates that MIN captures quantum correlations that are more general than entanglement and are not completely destroyed by thermal noise. From a physical perspective, this robustness arises because MIN quantifies the maximal global disturbance caused by local measurements on one subsystem, rather than the nonseparability of the state itself. Even when the detectors' joint state becomes separable, local measurements on one detector can still induce a nontrivial global change in the joint density matrix, signifying the presence of measurement-induced quantum correlations. Therefore, the nonzero asymptotic value of MIN at $T\to\infty$ reflects the residual quantumness of the accelerated detectors interacting with the field vacuum.

\section{Conclusion}\label{Conclusion}
In this work, we have investigated measurement-induced nonlocality in two uniformly accelerating Unruh--DeWitt detectors interacting with a massless scalar field. We find that the response of MIN to the Unruh temperature depends strongly on the initial correlations. Depending on the initial state, MIN can exhibit monotonic suppression, suppression followed by recovery, or monotonic enhancement. Moreover, MIN remains finite in the high-temperature limit, showing that strong Unruh thermalization does not necessarily erase this form of nonclassical correlation. This asymptotic persistence contrasts with the vanishing MIN previously reported for bosonic field modes in the infinite-acceleration limit~\cite{Tian2013}. Our results therefore indicate that the effect of acceleration on MIN is not universally destructive, but is determined by the interplay between Unruh thermalization and the initial correlation structure. It would be interesting to examine whether these behaviors persist for more general detector configurations and alternative measures of measurement-induced quantum correlations.

\begin{acknowledgments}
This work was supported by the National Natural Science Foundation of China (Grant Nos. 12175106 and 92365110), the Natural Science Foundation of Jiangsu Province, China (Grant No. BK20240612), the Natural Science Research Start-up Foundation of Recruiting Talents of Nanjing University of Posts and Telecommunications (Grant No. NY222123), and the Natural Science Foundation of Nanjing University of Posts and Telecommunications (Grant No. NY223069).
\end{acknowledgments}

\bibliography{ref}

\end{document}